\documentclass[prl,aps,secnumarabic,showpacs,twocolumn]{revtex4}
\usepackage{bm,latexsym,amsmath,amsfonts,graphicx,axodraw,subfigure,color}
\usepackage{hyperref,url,hhline}

\allowdisplaybreaks[4]

\begin{document}

%\fontsize{12pt}{14pt}\selectfont

\title{$\mathbb{Z}_2$ Symmetry Prediction for the Leptonic Dirac {\tt CP} Phase}
\author{
Shao-Feng Ge$^{1,}$\footnote{Electronic address: gesf02@gmail.com}, 
Duane A. Dicus$^{2,}$\footnote{Electronic address: dicus@physics.utexas.edu}, 
and Wayne W. Repko$^{3,}$\footnote{Electronic address: repko@pa.msu.edu}}
\affiliation{
$^1$Center for High Energy Physics, Tsinghua University, Beijing 100084, China  \\
$^2$Physics Department, University of Texas, Austin, TX 78712  \\
$^3$Department of Physics and Astronomy, Michigan State University, East Lansing MI 48824}

\date{\today}

\begin{abstract}
Model-independent consequences of applying a generalized hidden horizontal $\mathbb Z_2$ 
symmetry to the neutrino mass matrix are explored. The Dirac 
{\tt CP} phase $\delta_D$ can be expressed in terms of the three mixing angles as 
$4 c_a s_a c_s s_s s_x \cos \delta_D = (s^2_s - c^2_s s^2_x) (c^2_a - s^2_a)$ where the 
$s_i,\,c_i$ are sines and cosines of the atmospheric, solar, and reactor angles.
This relation is independent of neutrino masses and whether neutrinos are Dirac- or Majorana-type.
Given the present constraints on the angles, $\delta_D$ is constrained to be almost maximal, 
a result which can be explored in experiments such as NO$\nu$A and T2K. The Majorana {\tt CP} 
phases do not receive any constraint and are thus model-dependent.
Also a distribution of $\theta_x$ with a lower limit is obtained without specifying
$\delta_D$.
\end{abstract}
\pacs{14.60.Pq \hfill Phys. Let. B {\bf 702}, 220-223 (2011)}

\maketitle

{\it Introduction} --
The lepton sector has a quite different mixing pattern from that of the quarks. 
In the PMNS parameterization \cite{PMNS},
the atmospheric mixing angle $\theta_a \equiv \theta_{23}$ is almost maximal and the solar mixing 
angle $\theta_s \equiv \theta_{12}$ is also large. The reactor mixing angle $\theta_x \equiv \theta_{13}$ 
is small with an upper bound of about 10 degrees at the 1$\sigma$ level. The current global fit is summarized 
in Table~\ref{tab:data}.

\begin{table}[h]
\begin{center}
{\small
\begin{tabular}{|c||c|c|c|}
\hline
& & & \\[-3mm]
& $\sin^2\theta_s\,(\theta_s)$
& $\sin^2\theta_a\,(\theta_a)$ 
& $\sin^2\theta_x\,(\theta_x)$
\\
& & & \\[-3mm]
\hhline{|=::=|=|=|}
& & & \\[-4mm]
CV
& $0.312$~($34.0^\circ$) 
& $0.466$~($43.0^\circ$) 
& $0.016$~($7.3^\circ$)
\\[1mm]
\hline
& & & \\[-3mm]
  $1\sigma$ Range 
& $0.294$-$0.331$ 
& $0.408$-$0.539$ 
& $0.006$-$0.026$
\\[0.6mm]
& ($32.8$-$35.1^\circ$) 
& ($39.7$-$47.2^\circ$) 
&  ($4.4$-$9.3^\circ$)
\\[1mm]
\hline
\end{tabular}
} 
\caption{The global fit for the neutrino mixing angles \cite{Fogli:2008jx}.
The first row gives the central values.}
\label{tab:data}
\end{center}
\end{table}

Upcoming neutrino experiments (Daya Bay \cite{DayaBay}, Double CHOOZ \cite{DoubleCHOOZ}, and Reno \cite{reno}) 
will make precise measurements of the three mixing angles, especially $\theta_x$, and the Dirac {\tt CP} 
phase $\delta_D$ which will be indirectly measured by experiments such as NO$\nu$A \cite{nova} and 
T2K \cite{t2k}. A sizeable $\theta_x$ \cite{Fogli:2008jx,GonzalezGarcia:2010er,Mezzetto:2010zi} is 
crucial for pinning down $\delta_D$ since they always appear as the product $\sin\theta_x\,e^{i\delta_D}$. 

A model independent sign from experiment is $\mu$--$\tau$ symmetry \cite{MuTau} in the diagonal basis 
of the charged leptons \cite{DGR}. Under $\mu$--$\tau$ symmetry $\theta_x$ vanishes and $\theta_a$ is maximal.
This is incorporated in the widely accepted {\it tribimaximal mixing} \cite{tribimaximal}.

The essential point of $\mu$--$\tau$ symmetry is that it is a {\it residual symmetry} which directly determines the mixing 
pattern. Nevertheless, having one residual symmetry is not sufficient since $\mu$--$\tau$ symmetry 
can determine just two of the three mixing angles. Another $\mathbb Z^s_2$ symmetry, 
which determines the solar mixing angle, has been proposed in \cite{Lam:2008sh}.

However, deviations from 
the tribimaximal pattern are still allowed and need not be small 
\cite{Fogli:2008jx, Altarelli:2010gt, tribi-or-not}.
First, it is not accurate enough, 
especially for $\theta_x$. It can serve as a zeroth-order approximation but should receive higher order 
corrections. Also, if tribimaximal mixing were exact, there would be no effect of the Dirac {\tt CP} phase.
This is not what we would prefer, especially from the perspective of leptogenesis. 
Thus, $\mu$--$\tau$ symmetry should be abandoned.
Secondly, the solar mixing angle deviates from the tribimaximal one by 
more than one degree. A generalization 
with $\theta_s$ being set free is explored in \cite{GHY,DGH}. Also, in one model, $\theta_s$ is expressed in terms 
of a golden ratio \cite{golden} while another scheme is realized with dodeca-symmetry \cite{Kim:2010zub}.
We will explore the model-independent consequences of this generalized $\mathbb Z^s_2$ symmetry 
without assuming $\mu$--$\tau$ symmetry.

{\it Symmetry and Mixing} --
The neutrino mixing matrix can be determined by two $\mathbb Z_2$ symmetries \cite{Lam:2008sh}
of which the solar mixing angle $\theta_s$ is constrained by
\begin{equation}
  G_1(k)
=
  \frac{1}{2+k^2}
  \left\lgroup
  \begin{matrix}
    2-k^2 & 2k & 2k  \\
    2k & k^2 & -2  \\
    2k & -2 & k^2
  \end{matrix}
  \right\rgroup \,,
\label{G1} 
\end{equation}
the generator of the generalized $\mathbb Z^s_2$ symmetry \cite{GHY}.
Tribimaximal mixing corresponds to $k=-2$. Another choice is $k = - 3 / \sqrt 2$
with $\theta_s \approx 33.7^\circ$ which fits the data better.
The following discussion will concentrate on imposing only $G_1$.

In general, there are three possible horizontal symmetries \cite{Lam:2008sh}, $G_i$, $i=1,2,3$, 
which satisfy $G_i U_\nu = U_\nu d^{(i)}_\nu$ where $U_{\nu}$ is the mixing matrix of the neutrino mass 
matrix and the $d^{(i)}_\nu$ are diagonal rephasing matrices. We can write this relation in two equivalent 
forms,
\begin{equation}
  U^\dagger_\nu G_i U_\nu
=
  d^{(i)}_\nu
\qquad \Leftrightarrow \qquad
  G_i
=
  U_\nu d^{(i)}_\nu U^\dagger_\nu \,.
\label{eq:UGU}
\end{equation}
For Majorana neutrinos, the diagonal elements of $d^{(i)}_\nu$ are $\pm 1$ and thus there are only eight 
possibilities for $d^{(i)}_\nu$ spanning a product group $\mathbb Z_2 \otimes \mathbb Z_2 \otimes \mathbb Z_2$. 
But only $\mathbb Z_2 \otimes \mathbb Z_2$ is effective because the third $\mathbb Z_2$ just contributes 
an overall $+1$ or $-1$ factor. Since the $G_i$ are similarity transformed representations of the corresponding 
$d^{(i)}_\nu$ through $U_\nu$, they are equivalent to $d^{(i)}_\nu$. In other words, {\it the effective residual 
symmetry of lepton mixing is $\mathbb Z_2 \otimes \mathbb Z_2$ for Majorana neutrinos}. We call $d^{(i)}_\nu$ 
the {\bf kernel} of $G^{(i)}_\nu$.

From the first form of (\ref{eq:UGU}) we see that $U_\nu$ is a diagonalization 
matrix of $G_i$ with corresponding diagonalized matrix $d^{(i)}_\nu$. Given a
group representation, the mixing matrix can be obtained
by diagonalizing the representation matrix without resorting to 
the mass matrix. This provides a very convenient and direct
way of determining the mixing matrix. As no explicit mass matrix, and hence no mass 
eigenvalues, are involved, the relation between the horizontal symmetry and
the mixing matrix is mass-independent \cite{Lam:2006wm}. Also, the second
form tells us that the residual symmetries can be constructed in terms of the 
mixing matrix. In this sense, the symmetry can be determined phenomenologically.

For the generalized $G_1(k)$, with $d^{(1)}_\nu = \mbox{diag}(-1,1,1)$, 
there is a special form of the mixing matrix, 
\begin{equation}
  U_k
\equiv
  \left\lgroup
  \begin{matrix}
    \frac {-k}{\sqrt{2 + k^2}} & \frac {- \sqrt 2}{\sqrt{2 + k^2}} & 0 \\
    \frac 1 {\sqrt{2 + k^2}} & \frac {- k} {\sqrt{2 (2 + k^2)}} & - \frac 1 {\sqrt 2} \\
    \frac 1 {\sqrt{2 + k^2}} & \frac {- k} {\sqrt{2 (2 + k^2)}} &   \frac 1 {\sqrt 2}
  \end{matrix}
  \right\rgroup \,,
  \label{eq:Uk}
\end{equation}
with maximal $\theta_a$ and vanishing $\theta_x$.

{\it Reconstruction of Mixing Matrix} --
Since there is a degeneracy between the eigenvalues of $G_1$, its diagonalization
matrix is not unique. From (\ref{eq:UGU}) and (\ref{eq:Uk}) we get,
\begin{equation}
  G_1
=
  U_\nu d^{(1)}_\nu U^\dagger_\nu
=
  U_k U_T d^{(1)}_\nu U^\dagger_T U^\dagger_k \,,
  \label{eq:G1U}
\end{equation}
where $U_\nu$ denotes the physical neutrino mixing matrix.
Thus the physical neutrino mixing matrix $U_\nu$ can be expressed as
%\begin{equation}
$
  U_\nu
\equiv
  U_k U_T
$.
%\label{eq:Unuk}
%\end{equation}
%
The freedom of rotating between the degenerate eigenstates is represented
by a 2--3 unitary rotation parameterized as,
\begin{equation}
  U_T
\equiv
  \left\lgroup
  \begin{matrix}
    1 \\
  & c_T & - s_T e^{i \beta_4} \\
  & s_T e^{- i \beta_4} &   c_T
  \end{matrix}
  \right\rgroup
  \left\lgroup
  \begin{matrix}
    e^{i \beta_1} \\
  & e^{i \beta_2} \\
  & & e^{i \beta_3}
  \end{matrix}
  \right\rgroup \,,
  \label{eq:UT}
\end{equation}
with $c_T \equiv \cos \theta_T$ and $s_T \equiv \sin \theta_T$.
The diagonal rephasing matrix $d^{(1)}_\nu$ is invariant under $U_T$, namely
$d^{(1)}_\nu = U_T d^{(1)}_\nu U^\dagger_T$.

Also, the physical mixing matrix $U_\nu$ can be generally parameterized as
$\mathcal P_\nu \mathcal U_\nu \mathcal Q_\nu$ where $\mathcal U_\nu$
is the standard parametrization of MNS matrix, 
\begin{eqnarray}
 \mathcal U_{\nu}
= \hspace{-1mm}
   \left\lgroup
   \begin{matrix}
   c_sc_x & -s_sc_x & -s_xe^{i\delta_D} \\
   s_sc_a-c_ss_as_xe^{-i\delta_D} & c_sc_a+s_ss_as_xe^{-i\delta_D} & -s_ac_x \\
   s_ss_a+c_sc_as_xe^{-i\delta_D} & c_ss_a-s_sc_as_xe^{-i\delta_D} & c_ac_x
   \end{matrix}
  \right\rgroup \hspace{-1mm} \,,
\end{eqnarray}
while
$\mathcal P_\nu \equiv \mbox{diag}(e^{i \alpha_1}, e^{i \alpha_2}, e^{i \alpha_3})$
and $\mathcal Q_\nu \equiv \mbox{diag}(e^{i \phi_1}, e^{i \phi_2}, e^{i \phi_3})$
are two diagonal rephasing matrices \cite{GHY}.
The phases $\phi_i$ in $\mathcal Q_\nu$ are Majorana {\tt CP} phases 
while $\mathcal P_\nu$ is a diagonal rephasing matrix which does not have any 
direct physical significance.

These two expressions for the physical mixing matrix $U_\nu$ must be equal,
\begin{equation}
  U_k U_T
=
  \mathcal P_\nu \mathcal U_\nu \mathcal Q_\nu \,.
\label{eq:mUk}
\end{equation}
The Majorana {\tt CP} phases $\phi_i$ in $\mathcal Q_\nu$ could be absorbed 
by the phases $\beta_i (i = 1,2,3)$ of $U_T$.

{\it Dirac CP Phase and Mixing Angles} --
We will first compare the elements of the first row of the two matrices on either side of Eq.(\ref{eq:mUk}).
They give
\begin{subequations}
\begin{eqnarray}
  \phi_1 - \beta_1
=
- \alpha_1 \,,
& &
  c_s c_x 
=
  \frac {- k} {\sqrt{2 + k^2}} \,,
\qquad
\label{eq:phi-beta-1}
\\
  \phi_2 - \beta_2
=
- \alpha_1 \,,
& &
  s_s c_x
=
  \frac {\sqrt 2 c_T}{\sqrt{2 + k^2}} \,, 
\label{eq:phi-beta-2}
\\
  \phi_3 - \beta_3
=
  \beta_4 - \delta_D - \alpha_1 \,,
& &
  s_x
=
  \frac {- \sqrt 2 s_T}{\sqrt{2 + k^2}} \,.
\label{eq:phi-beta-3}
\end{eqnarray}
\label{eq:phi-beta}
\end{subequations}
The matching of overall {\tt CP} phases leaves some freedom which gives, at most,  minus signs.
These signs will not affect the final physical result so, for simplicity,  we will omit them.
From (\ref{eq:phi-beta}) we can see that only the differences $\phi_i - \beta_i$ are relevant.
Majorana phases $\phi_i$ can not be uniquely determined.
The mixing angles, $\theta_x$ and $\theta_s$ can be expressed as functions of $k$ and 
$\theta_T$. Of the three relations for the mixing angles only two are independent
because of unitarity. Conversely, $k$ and $\theta_T$ can be expressed as functions of
mixing angles,
\begin{eqnarray}
  k^2
=
  \frac {2 c^2_s c^2_x}
        {1 - c^2_s c^2_x} \,,
\qquad
  s^2_T
=
  \frac {s^2_x}
        {1 - c^2_s c^2_x} \,.
\label{eq:kTD}
\end{eqnarray}
From (\ref{eq:kTD}) we can estimate the approximate value of the fitting parameters. 
Since $s^2_x$ is tiny, so is $s^2_T$, and $k^2 \approx 2 \cot^2 \theta_s$.
We see that the value of $k$ is close to $2$ but with some deviation. According to 
(\ref{eq:kTD}), $s^2_T$ is approximately $3 \, s^2_x$. The central value of $\theta_x$ we 
used is about $7.3^\circ$ rendering $\theta_T \approx 13^\circ$.

The $(21)$ and $(31)$ elements of Eq.(\ref{eq:mUk}) give the 
differences between $\alpha_i$ in terms of mixing angles and $\delta_D$,
\begin{subequations}
\begin{eqnarray}
  e^{i (\alpha_2 - \alpha_1)}
& = &
  \frac 1 { \sqrt{2 + k^2} \left( s_s c_a - c_s s_a s_x e^{- i \delta_D} \right)} \,,
\label{eq:alpha-21-31-a}
\\
  e^{i (\alpha_3 - \alpha_1)}
& = &
  \frac 1 { \sqrt{2 + k^2} \left( s_s s_a + c_s c_a s_x e^{- i \delta_D} \right)} \,.
\qquad
\label{eq:alpha-21-31}
\end{eqnarray}
\end{subequations}
The sum of the $(23)$ and $(33)$ elements gives,
\begin{eqnarray}
  \left[ e^{i \delta_3} c_a - e^{i \delta_2} s_a \right] c_x
=
  \sqrt{\frac{2 k^2}{2 + k^2}} s_T e^{i \beta_4} \,,
  \label{eq:elements-33}
\end{eqnarray}
with $\delta_i \equiv \alpha_i - \alpha_1 + \beta_4 - \delta_D$. The common $\beta_4$
cancels.

By comparing (\ref{eq:elements-33}) with (\ref{eq:phi-beta-3}) we can eliminate
$\theta_T$. Then (\ref{eq:alpha-21-31-a}) and   (\ref{eq:alpha-21-31}) can be used to remove the Majorana phases $\alpha_i$.
The parameter $k$ can be expressed in terms of the mixing angles through (\ref{eq:phi-beta-1}).
When all of these are combined a relation between
%Dirac {\tt CP} phase 
$\delta_D$ and the three mixing angles emerges,
\begin{equation}
  \cos \delta_D
=
  \frac {(s^2_s - c^2_s s^2_x)(c^2_a - s^2_a)}
        {4 c_a s_a c_s s_s s_x} \,.
\label{eq:G-relation-a}
\end{equation}
Note that this correlation between the mixing angles and $\delta_D$ is 
independent of unphysical parameters and thus it can be used to compare 
with experimental results directly. With vanishing $\delta_D$ 
(\ref{eq:G-relation-a}) will reduce to the result of \cite{Dicus:2010yu}.

\begin{figure}[h]
\centering
\subfigure[\, $\theta_x$ Histogram]
          {\label{fig:thetax-x} \includegraphics[width=8cm,height=5cm]{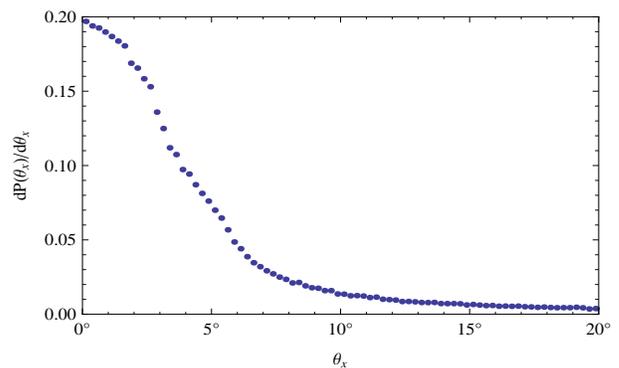}}
\qquad
\subfigure[\, $\theta_x$ v.s. $\theta_a$]
          {\label{fig:thetax-xa} \includegraphics[width=8cm,height=5.1cm]{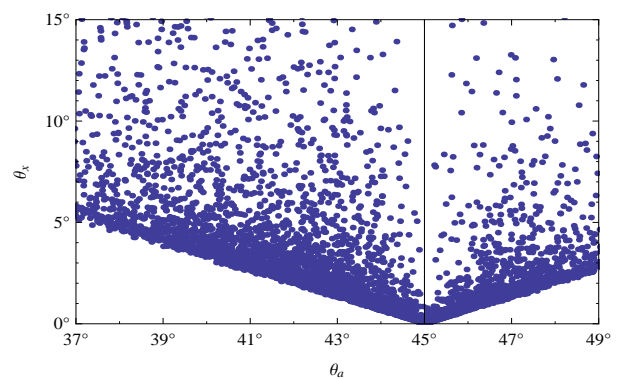}}
\caption{Predictions of $\theta_x$ in terms of $\delta_D, \theta_a$, and $\theta_s$ at the 90\% C.L.}
\label{fig:thetax}
\end{figure}

As $\theta_x$ is small and $\theta_a$ is nearly maximal, we can define 
$\theta_x \equiv 0^\circ + \delta_x$ and $\theta_a \equiv 45^\circ + \delta_a$. 
To leading order (\ref{eq:G-relation-a}) reduces to,
\begin{equation}
  \frac {\delta_x}{\delta_a}
=
- \frac {\tan \theta_s}{\cos \delta_D} \,,
\label{eq:dx-da-D}
\end{equation}
which is the main result of \cite{GHY} obtained from the minimal seesaw model. 
In that model, both $\mu$--$\tau$ and $G_1$ 
symmetries are imposed at leading order and a soft mass term is used to break $\mu$--$\tau$
without affecting $G_1$. This is similar to our treatment here where only $G_1$ is applied 
to constrain the mixing angles and the Dirac {\tt CP} phase without involving $\mu$--$\tau$
symmetry. The difference is that an expansion method was used in \cite{GHY} while
the discussion in the current work is more general and, most importantly,
model-independent.

{\it Phenomenological Consequences} --
The relation (\ref{eq:G-relation-a}) can be used to put limits on $\theta_x$ from,
\begin{equation}
  \sin \theta_x
=
  \left[
   \mp \sqrt{c^2_D + \cot^2 2 \theta_a}
  - c_D
  \right] \tan 2 \theta_a \tan \theta_s\,,
\end{equation}
with $c_D \equiv \cos \delta_D$.
The result for the lower sign is shown in Fig.\,\ref{fig:thetax-x} where 
$\delta_D$ is assumed to be uniformly distributed in the range  $(0,2 \pi)$ 
and normal distributions for $\theta_a$ and $\theta_s$ are used with central 
values and standard deviations given by Table \ref{tab:data}. Alternately we 
can use the limits $1\,\ge\,\cos\delta_D\,\ge\,-1$ to get limits on $s_x$,
\begin{equation}\label{limits}
  \frac{s_s}{c_s} \frac{c_a+s_a}{|c_a-s_a|}
\,\ge\,
  \sin\theta_x
\,\ge\,
  \frac{s_s}{c_s}\frac{|c_a-s_a|}{c_a+s_a}\,.
\end{equation}
The upper limit is of no use but the lower limit is clearly shown in Fig.\,\ref{fig:thetax-xa}.

\begin{figure}
\centering
\subfigure[\, Dirac CP Phase $\delta_D$]
          {\label{fig:DJ-D} \includegraphics[width=8cm,height=5cm]{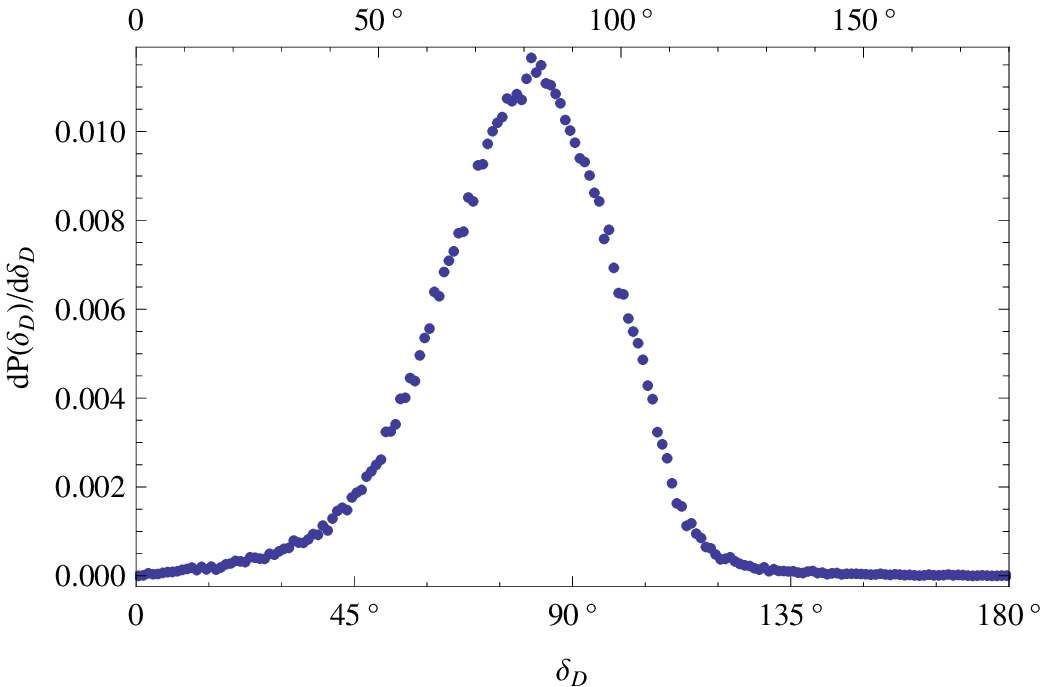}}
\qquad
\subfigure[\, Jarlskog Invariant $J_\nu$]
          {\label{fig:DJ-J} \includegraphics[width=8cm,height=4.7cm]{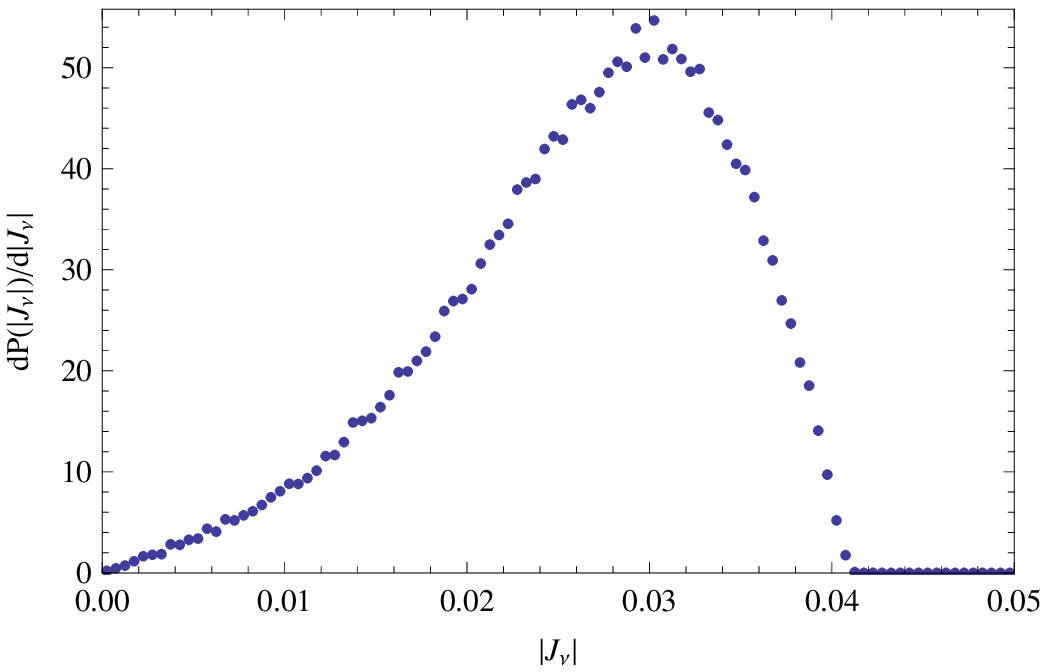}}
\caption{Normalized differential probability distribution of (a) the Dirac {\tt CP} phase $\delta_D$
         and (b) Jarlskog Invariant $J_\nu$ at 90\% C.L. The distribution of $\delta_D$ is symmetric 
         with respect to $0^0$ so there would be another peak near $-90^0$. The distribution of $J_\nu$
         is also symmetric around $0$.}
\label{fig:DJ}
\end{figure}

Since the three mixing angles have been measured, a prediction of $\delta_D$ 
can be obtained using (\ref{eq:G-relation-a}). This is plotted in Fig.\,\ref{fig:DJ-D} where 
there exists a corresponding distribution in the range of $(-180^0,0^0)$.
We can see that $\delta_D$ is constrained 
to be almost maximal which is a direct consequence of the fact that $\theta_a$ is closer to its 
zeroth-order approximation than $\theta_x$. From (\ref{eq:G-relation-a}) we see that if $\delta_a$ 
vanishes, namely $c_a = s_a$, $\cos \delta_D$ would also vanish.
A similar result is obtained in \cite{Antusch:2011sx} in a model-dependent way.

Experimentally,  $\delta_D$ is measured through the {Jarlskog invariant} \cite{Jarlskog:1985ht} 
$J_\nu \equiv c_a s_a c_s s_s c^2_x s_x \sin \delta_D $ whose distribution is shown in 
Fig.\,\ref{fig:DJ-J}.  
There is an explicit upper limit around $0.04$. 
The distribution peaks at $|J_\nu| \approx 0.03$ which can be 
estimated with the best fit values listed in Table\,\ref{tab:data}. This will be tested 
by future experiments \cite{nova,t2k}.

{\it Generalization to Dirac Type Neutrino} --
The direct relation between $G_i$ and $U_\nu$ shown in (\ref{eq:UGU}) is the same 
for Dirac- and Majorana-type
fermions \cite{Lam:2008sh}. The only difference comes from the kernel $d^{(i)}_\nu$.  
For Majorana neutrinos $d^{(i)}_\nu d^{(i)}_\nu = I$, while 
for Dirac neutrinos the kernel can be complex, $d^{(i)}_\nu d^{(i) \dagger}_\nu = I$.

As long as $G_1$ is the same, its kernel $d^{(1)}_\nu$ is still $(-1,1,1)$ for either 
Majorana or Dirac neutrinos. Thus, all the above results also apply for 
Dirac neutrinos. This result is quite general and  has not been noticed before. 
In \cite{Lam:2008sh} the author considered the group $S_4$ generated by subgroups 
$\mathcal F$ and $\mathcal G$ for the charged lepton and neutrino sectors respectively 
with the neutrinos being Majorana-type. The element $G_i$ of the subgroup $\mathcal G$ 
has a kernel with diagonal elements being $-1\,,+1\,,+1$ in some order. This 
satisfies the constraint on not only Majorana- but also Dirac-type kernels.  
Thus $\mathcal G$ is also true for the case of Dirac-type neutrinos.
The generated $S_4$ symmetry applies to both Dirac- and Majorana-type 
neutrinos, not just the Majorana-type as discussed in \cite{Lam:2008sh}.

{\it Conclusions} -- 
Model-independent consequences of a generalized $G_1$ symmetry are explored. Due to 
degeneracy between the eigenvalues of $G_1$, the mixing matrix cannot be uniquely 
determined. Nevertheless, $G_1$ invariance gives a relation (\ref{eq:G-relation-a})
between the mixing angles and the Dirac {\tt CP} phase $\delta_D$ for both Majorana 
and Dirac neutrinos. This can be used to predict $\delta_D$, and consequently the Jarlskog 
invariant $J_\nu$, in terms of the measured mixing angles leading to an almost maximal 
$\delta_D$ as shown in Fig.\,\ref{fig:DJ}. 
This appears to be the first time that a direct 
relation between a horizontal symmetry and the leptonic Dirac {\tt CP} phase 
has been established in a model-independent way.  These results will be 
tested by next generation of neutrino experiments 
\cite{DayaBay,DoubleCHOOZ,reno,nova,t2k}. Comparison between this prediction 
of $\delta_D$ and experiment can tell us whether or not there is a horizontal 
symmetry and, if so, what type of symmetry it is. Since the prediction is 
independent of model assignments, its conformation should be robust and conclusive.
The Majorana {\tt CP} phases are not constrained;  they are model-dependent and 
must be studied case by case. In addition a clear lower limit on $\theta_x$ is 
obtained as shown in Fig.\,\ref{fig:thetax}.

{\it Acknowledgments} --
SFG is grateful to Hong-Jian He for useful discussions and kind support.
DAD was supported in part by the U. S. Department of Energy under grant No. DE-FG03-93ER40757. 
WWR was supported in part by the National Science Foundation under Grant PHY-0555544.


\begin{thebibliography}{99}

%\bibitem{19}B. Pontecorvo, Sov.Phys.JETP{\bf 6},429(1958); Z. Maki, M. Nakagawa, S. Sakata, Prog. Theor.Phys.{\bf 28},870(1962).
\bibitem{PMNS}
  B.~Pontecorvo,
% ``{\it Mesonium and antimesonium},''
  Sov.\ Phys.\ JETP {\bf 6}, 429 (1957)
  [Zh.\ Eksp.\ Teor.\ Fiz.\  {\bf 33}, 549 (1957)];
  Z.~Maki, M.~Nakagawa and S.~Sakata,
% ``{\it Remarks on the unified model of elementary particles},''
  Prog.\ Theor.\ Phys.\  {\bf 28}, 870 (1962).


%\cite{Fogli:2008jx}
\bibitem{Fogli:2008jx}
  G.~L.~Fogli, E.~Lisi, A.~Marrone, A.~Palazzo and A.~M.~Rotunno,
% ``{\it Hints of $\theta_{13} > 0$ from global neutrino data analysis},''
  Phys.\ Rev.\ Lett.\  {\bf 101}, 141801 (2008)
  [arXiv:0806.2649 [hep-ph]].
  %%CITATION = PRLTA,101,141801;%%



\bibitem{DayaBay}
  X.~Guo {\it et al.}  [Daya-Bay Collaboration],
% ``{\it A precision measurement of the neutrino mixing angle $\theta_{13}$ using
% reactor antineutrinos at Daya Bay},''
  arXiv:hep-ex/0701029;
%
  M.~C.~Chu  [Daya Bay Collaboration],
% ``{\it Precise measurement of $\theta_{13}$ at Daya Bay},''
  arXiv:0810.0807 [hep-ex];
%
  W.~Wang  [Daya Bay Collaboration],
% ``{\it The hunt for $\theta_{13}$ at the Daya Bay nuclear power plant},''
  AIP Conf.\ Proc.\  {\bf 1222}, 494 (2010)
  [arXiv:0910.4605 [hep-ex]].


\bibitem{DoubleCHOOZ}
  F.~Ardellier {\it et al.}  [Double Chooz Collaboration],
% ``{\it Double Chooz: A search for the neutrino mixing angle $\theta_{13}$},''
  arXiv:hep-ex/0606025.


\bibitem{reno}
  J.~K.~Ahn {\it et al.}  [RENO Collaboration],
% ``{\it RENO: An Experiment for Neutrino Oscillation Parameter $\theta_{13}$ Using
% Reactor Neutrinos at Yonggwang},''
  arXiv:1003.1391 [hep-ex].

\bibitem{nova}
  D.~S.~Ayres {\it et al.}  [NO$\nu$A Collaboration],
% ``{\it NO$\nu$A proposal to build a 30-kiloton off-axis detector to study neutrino
% oscillations in the Fermilab NuMI beamline},''
  arXiv:hep-ex/0503053;
%
  O.~Mena, S.~Palomares-Ruiz and S.~Pascoli,
% ``{\it Determining the neutrino mass hierarchy and {\tt CP} violation in NO$\nu$A with  a
% second off-axis detector},''
  Phys.\ Rev.\  D {\bf 73}, 073007 (2006)
  [arXiv:hep-ph/0510182];
%
  D.~S.~Ayres {\it et al.}  [NO$\nu$A Collaboration],
 ``{\it The NO$\nu$A Technical Design Report}.''
%
%  R.~Plunkett  [NOvA Collaboration],
%  ``{\it Status of the NOvA experiment},''
%  J.\ Phys.\ Conf.\ Ser.\  {\bf 120}, 052044 (2008).


\bibitem{t2k}
  Y.~Itow {\it et al.}  [The T2K Collaboration],
% ``{\it The JHF-Kamioka neutrino project},''
  arXiv:hep-ex/0106019.
%
%  P.~Huber, M.~Maltoni and T.~Schwetz,
%  ``{\it Resolving parameter degeneracies in long-baseline experiments by
%  atmospheric neutrino data},''
%  Phys.\ Rev.\  D {\bf 71}, 053006 (2005)
%  [arXiv:hep-ph/0501037].


%\cite{GonzalezGarcia:2010er}
\bibitem{GonzalezGarcia:2010er}
  M.~C.~Gonzalez-Garcia, M.~Maltoni and J.~Salvado,
% ``{\it Updated global fit to three neutrino mixing: status of the hints of 
%    $\theta_{13} > 0$},''
  JHEP {\bf 1004}, 056 (2010)
  [arXiv:1001.4524 [hep-ph]].
  %%CITATION = JHEPA,1004,056;%%


%\cite{Mezzetto:2010zi}
\bibitem{Mezzetto:2010zi}
  M.~Mezzetto and T.~Schwetz,
% ``{\it $\theta_{13}$: phenomenology, present status and prospect},''
  J.\ Phys.\ G {\bf 37}, 103001 (2010)
  [arXiv:1003.5800 [hep-ph]].


\bibitem{MuTau}
  R.~N.~Mohapatra and S.~Nussinov,
% ``{\it Bimaximal neutrino mixing and neutrino mass matrix},''
  Phys.\ Rev.\  D {\bf 60}, 013002 (1999)
  [arXiv:hep-ph/9809415];
%
  C.~S.~Lam,
% ``{\it A 2-3 symmetry in neutrino oscillations},''
  Phys.\ Lett.\  B {\bf 507}, 214 (2001)
  [arXiv:hep-ph/0104116];
%
%  C.~S.~Lam,
%  ``{\it Neutrino 2-3 symmetry and inverted hierarchy},''
%  Phys.\ Rev.\  D {\bf 71}, 093001 (2005)
%  [arXiv:hep-ph/0503159].
%
  T.~Fukuyama and H.~Nishiura,
  %``Mass matrix of Majorana neutrinos,''
  arXiv:hep-ph/9702253.
  %%CITATION = HEP-PH/9702253;%%



\bibitem{DGR}
D.~A.~Dicus, S.-F.~Ge and W.~W.~Repko, Phys. Rev. D {\bf 82}, 033005 (2010)
% ``{\it Neutrino mixing with broken $S_3$ symmetry},''
  arXiv:1004.3266 [hep-ph].
  %%CITATION = ARXIV:1004.3266;%%



\bibitem{tribimaximal}
  P.~F.~Harrison, D.~H.~Perkins and W.~G.~Scott,
% ``{\it Tri-bimaximal mixing and the neutrino oscillation data},''
  Phys.\ Lett.\  B {\bf 530}, 167 (2002)
  [arXiv:hep-ph/0202074]; 
%
  Z.~Z.~Xing,
% ``{\it Nearly tri-bimaximal neutrino mixing and CP violation},''
  Phys.\ Lett.\  B {\bf 533}, 85 (2002)
  [arXiv:hep-ph/0204049];
%
  X.~G.~He and A.~Zee,
% ``{\it Some simple mixing and mass matrices for neutrinos},''
  Phys.\ Lett.\  B {\bf 560}, 87 (2003)
  [arXiv:hep-ph/0301092].


%\cite{Lam:2008sh}
\bibitem{Lam:2008sh}
  C.~S.~Lam,
% ``{\it The Unique Horizontal Symmetry of Leptons},''
  Phys.\ Rev.\  D {\bf 78}, 073015 (2008)
  [arXiv:0809.1185 [hep-ph]].




%\cite{Altarelli:2010gt}
\bibitem{Altarelli:2010gt}
  G.~Altarelli and F.~Feruglio,
% ``{\it Discrete Flavor Symmetries and Models of Neutrino Mixing},''
  Rev.\ Mod.\ Phys.\  {\bf 82}, 2701 (2010)
  [arXiv:1002.0211 [hep-ph]].
  %%CITATION = RMPHA,82,2701;%%

\bibitem{tribi-or-not}
  C.~H.~Albright and W.~Rodejohann,
% ``{\it Model-Independent Analysis of Tri-bimaximal Mixing: A Softly-Broken Hidden
% or an Accidental Symmetry?},''
  Phys.\ Lett.\  B {\bf 665}, 378 (2008)
  [arXiv:0804.4581 [hep-ph]];
%
  M.~Abbas and A.~Y.~Smirnov,
% ``{\it Is the tri-bimaximal mixing accidental?},''
  Phys.\ Rev.\  D {\bf 82}, 013008 (2010)
  [arXiv:1004.0099 [hep-ph]];
%
  C.~H.~Albright, A.~Dueck and W.~Rodejohann,
% ``{\it Possible Alternatives to Tri-bimaximal Mixing},''
  Eur.\ Phys.\ J.\  C {\bf 70}, 1099 (2010)
  [arXiv:1004.2798 [hep-ph]].



%\cite{Ge:2010js}
\bibitem{GHY}
  S.-F.~Ge, H.~J.~He and F.~R.~Yin,
% ``{\it Common Origin of Soft $\mu$-$\tau$ and {\tt CP} Breaking in Neutrino Seesaw and the
% Origin of Matter},''
  JCAP {\bf 1005}, 017 (2010)
  [arXiv:1001.0940 [hep-ph]].
  %%CITATION = JCAPA,1005,017;%%

\bibitem{DGH} D.A. Dicus, S.-F. Ge and H.-J. He, %''{\it Beyond Tribimaximal Mixing}", 
in preparation.

\bibitem{golden}
  A.~Datta, F.~S.~Ling and P.~Ramond,
% ``{\it Correlated hierarchy, Dirac masses and large mixing angles},''
  Nucl.\ Phys.\  B {\bf 671}, 383 (2003)
  [arXiv:hep-ph/0306002];
%
  Y.~Kajiyama, M.~Raidal and A.~Strumia,
% ``{\it The golden ratio prediction for the solar neutrino mixing},''
  Phys.\ Rev.\  D {\bf 76}, 117301 (2007)
  [arXiv:0705.4559 [hep-ph]];
%
  L.~L.~Everett and A.~J.~Stuart,
% ``{\it Icosahedral (A5) Family Symmetry and the Golden Ratio Prediction for Solar
% Neutrino Mixing},''
  Phys.\ Rev.\  D {\bf 79}, 085005 (2009)
  [arXiv:0812.1057 [hep-ph]];
%
  A.~Adulpravitchai, A.~Blum and W.~Rodejohann,
% ``{\it Golden Ratio Prediction for Solar Neutrino Mixing},''
  New J.\ Phys.\  {\bf 11}, 063026 (2009)
  [arXiv:0903.0531 [hep-ph]];
%
  F.~Feruglio and A.~Paris,
% ``{\it The Golden Ratio Prediction for the Solar Angle from a Natural Model with
% $A_5$ Flavour Symmetry},''
  arXiv:1101.0393 [hep-ph].


\bibitem{Kim:2010zub}
  J.~E.~Kim and M.~S.~Seo,
% ``{\it Quark and lepton mixing angles with a dodeca-symmetry},''
  arXiv:1005.4684 [hep-ph].






%\cite{Lam:2006wm}
\bibitem{Lam:2006wm}
  C.~S.~Lam,
% ``{\it Mass Independent Textures and Symmetry},''
  Phys.\ Rev.\  D {\bf 74}, 113004 (2006)
  [arXiv:hep-ph/0611017].
  %%CITATION = PHRVA,D74,113004;%%


%\cite{Dicus:2010yu}
\bibitem{Dicus:2010yu}
  D.~A.~Dicus, S.~F.~Ge and W.~W.~Repko, Phys. Rev. D {\bf 83}, 093007 (2011)
  %``Generalized Hidden $\mathcal{Z}_2$ Symmetry of Neutrino Mixing,''
  arXiv:1012.2571 [hep-ph].
  %%CITATION = ARXIV:1012.2571;%%



%\cite{Antusch:2011sx}
\bibitem{Antusch:2011sx}
  S.~Antusch, S.~F.~King, C.~Luhn and M.~Spinrath,
  %``Right unitarity triangles and tri-bimaximal mixing from discrete symmetries
  %and unification,''
  arXiv:1103.5930 [hep-ph].
  %%CITATION = ARXIV:1103.5930;%%



%\cite{Jarlskog:1985ht}
\bibitem{Jarlskog:1985ht}
  C.~Jarlskog,
%  ``{\it Commutator Of The Quark Mass Matrices In The Standard Electroweak Model And
%  A Measure Of Maximal {\tt CP} Violation},''
  Phys.\ Rev.\ Lett.\  {\bf 55}, 1039 (1985).



%\bibitem{GXZ}W-l. Guo, Z-z. Xing, and S. Zhou, Int.J.Mod.Phys. E{\bf 16}, 1 (2007).







%\bibitem{MuTauNu}
%  C.~S.~Lam,
%% ``{\it Neutrino 2-3 symmetry and inverted hierarchy},''
%  Phys.\ Rev.\  D {\bf 71}, 093001 (2005)
%  [arXiv:hep-ph/0503159].


%\cite{Friedberg:2010zt}
%\bibitem{Friedberg:2010zt}
%  R.~Friedberg and T.~D.~Lee,
%  ``{\it Deviations of the Lepton Mapping Matrix from the Harrison-Perkins-Scott
%  Form},''
%  arXiv:1008.0453 [hep-ph].
  %%CITATION = ARXIV:1008.0453;%%



%\cite{Kobayashi:1973fv}
%\bibitem{Kobayashi:1973fv}
%  M.~Kobayashi and T.~Maskawa,
%  ``{\it {\tt CP} Violation In The Renormalizable Theory Of Weak Interaction},''
%  Prog.\ Theor.\ Phys.\  {\bf 49}, 652 (1973).
  %%CITATION = PTPKA,49,652;%%

%\bibitem{MinimalSeesaw}
%  P.~H.~Frampton, S.~L.~Glashow and T.~Yanagida,
%  ``{\it Cosmological sign of neutrino {\tt CP} violation},''
%  Phys.\ Lett.\  B {\bf 548}, 119 (2002)
%  [arXiv:hep-ph/0208157];
%  M.~Raidal and A.~Strumia,
%  ``{\it Predictions of the most minimal see-saw model},''
%  Phys.\ Lett.\  B {\bf 553}, 72 (2003)
%  [arXiv:hep-ph/0210021].



%\bibitem{BDHL} 
%  V.~Barger, D.~A.~Dicus, H.~J.~He and T.~J.~Li,
%  ``{\it Structure of cosmological {\tt CP} violation via neutrino seesaw},''
%  Phys.\ Lett.\  B {\bf 583}, 173 (2004)
%  [arXiv:hep-ph/0310278].

%\bibitem{FLMPR}G.L. Fogli, {\it et. al.}, Phys. Rev. Lett. {\bf 101}, 141801 [arXiv:0806.2649] and arXiv:0809.2936[hep-ph].
%\bibitem{MINOS}P. Adamson {\it et. al.}, Phys. Rev. Lett. {\bf 103}, 261802-1 (2009).
%.\bibitem{STV}T.Schwetz, M.Tortola, J.W.F.Valle, New J. Phys. {\bf 10},113011 (2008).


\end{thebibliography}
\end{document}